\documentclass[11pt,twoside]{article}
\usepackage{asp2010}

\resetcounters

\begin{document}

\title{Residual Energy in Weak and Strong MHD Turbulence}
\author{Stanislav Boldyrev$^1$, Jean Carlos Perez$^{1,2}$, and Yuxuan Wang$^1$
\affil{$^{1}$Department of Physics, University of Wisconsin, 1150 University Ave, Madison, WI 53706, USA}
\affil{$^2$Space Science Center, University of New Hampshire, Durham, NH 03824, USA}
}

\begin{abstract}
Recent numerical and observational studies revealed that spectra of magnetic and velocity fluctuations in MHD turbulence have different scaling indexes. This intriguing feature has  been recently explained in the case of weak MHD turbulence, that is, turbulence consisting of weakly interacting Alfv\'en waves. However, astrophysical turbulence is strong in majority of cases. In the present work, we propose a unifying picture that allows one to address weak and strong MHD turbulence on the same footing.  We argue that magnetic and kinetic energies are different in both weak and strong MHD turbulence. Their difference, the so-called residual energy, is spontaneously generated by turbulence, it has the Fourier spectrum $E_r(k)=E_v(k)-E_b(k) \propto -f_w(k_\|/k_\perp) k_\perp^{-2}$ in weak turbulence, and $E_r(k)\propto -f_s(k_\|/k_\perp) k_\perp^{-3}$ in strong turbulence. Here $f_{w,s}(x)$ are functions declining fast for $x>C_{w,s}$ and not significantly varying for $x<C_{w,s}$ with some constants $C_{w,s}$, and $k_\|$ and $k_\perp$ the field-parallel and field-perpendicular wave vectors with respect to the applied strong uniform magnetic field. 
\end{abstract}

\section{Introduction}
Magnetic fields and turbulence are common in a variety of astrophysical plasmas, from planets and stars to interstellar and intergalactic media. Magnetic turbulence is also commonly invoked to explain small-scale features of the solar wind. Numerical simulations and analytic modeling play an important role in interpreting observational data. Recently, it has been found that magnetic and velocity fluctuations are not in equipartition in MHD turbulence \citep[e.g.,][]{podesta_rg07,tessein_etal09,chen11a,boldyrev_pbp11}, which seems to be at odds with basic assumptions of conventional models of MHD turbulence. The goal of the present contribution is to propose an explanation for this intriguing phenomenon.  
%While turbulence of ordinary nonmagnetized fluids has been well studied experimentally for decades, turbulence of magnetized fluids is relatively more difficult to study in laboratory. Therefore, numerical simulations and analytic modeling play an important role in developing the theory of MHD turbulence. 
%Recent high-resolution measurements of the solar wind fluctuations turn out to contradict some basic assumptions of conventional picture of MHD turbulence, which raised the questions from whether  the solar wind can be adequately described by MHD turbulence to whether the theory of MHD turbulence has been developed well enough to capture the observed effects. 
In contrast with ordinary incompressible turbulence, which is always in a strongly coupled state, incompressible MHD turbulence can exhibit two distinct regimes of weak and strong turbulence. This stems from the fact that the MHD system possesses Alfv\'en waves that can coalesce and scatter off each other. When during a singe interaction the wave amplitudes change only slightly, turbulence is weak, otherwise, it is strong. It is important to note however that even if MHD turbulence is weak at large scales, its strength increases toward small scales, so that the range of scales where weak MHD turbulence may exist is typically limited.   
%The original picture of MHD turbulence due to \cite{iroshnikov} and \cite{kraichnan65} suggested that MHD turbulence becomes asymptotically weak as the scale of the fluctuations decreases, thus ensuring analytical tractability of the phenomenon. However, the modern developmens suggest that the picture is, in fact, inverse: the strength of MHD turbulence increases at small scales. 
In this contribution we present a unifying model of MHD turbulence valid for both weak and strong regimes. The incompressible MHD equations for magnetic and velocity fields, ${\bf b}({\bf x}, t)$ and ${\bf v}({\bf x}, t)$, have especially useful form when written in the so-called Elsasser variables ${\bf z}^{\pm}={\bf
v}\pm{\bf b}$,
\begin{equation}
  \left( \frac{\partial}{\partial t}\mp{\bf v}_A\cdot\nabla\right){\bf 
  z}^\pm+\left({\bf z}^\mp\cdot\nabla\right){\bf z}^\pm = -\nabla P. 
  \label{mhd-elsasser}
\end{equation}
The equations are written in a frame with zero mean-flow velocity, ${\bf b}$ is
the fluctuating magnetic field normalized by $\sqrt{4 \pi \rho_0}$,
${\bf v}_A={\bf B}_0/\sqrt{4\pi \rho_0}$ is the Alfv\'en velocity corresponding to the
uniform magnetic field ${\bf B}_0$ (so that the total magnetic field is ${\bf B}={\bf B}_0+{\bf b}$),  $P=p/\rho_0+b^2/2$, it includes the
plasma pressure, $p$, and the magnetic pressure, $\rho_0$ is the constant 
mass density, and we neglect driving and dissipation terms  \citep[e.g.,][]{biskamp03,marsch_m87}. In what follows we assume that turbulence is driven at large scales, that can be  mimicked by adding forcing terms to the right-hand sides of Eqs.~(\ref{mhd-elsasser}). 
%The character of the forcing may vary in different systems -- one can drive the turbulence by exciting velocity fluctuation (the dynamo-type driving), by exciting Alfv\'en waves (${\bf z}^+$ and ${\bf z}^-$ fields), etc. In numerical simulations the driving is often optimized to address particular requirements, such as constant energy input per unit time, fixed amplitudes of large-scale fluctuations, specific correlations between magnetic and velocity fields, etc. However, 
Small-scale turbulence is expected to be independent of the large-scale driving  \citep[e.g.,][]{mason_cb08}. We will also assume that the uniform guide
field is strong compared to the rms fluctuations, that is, $b_{rms} \ll B_0$. 

\section{The energy spectrum}
The ideal MHD equations conserve the two Elsasser energies, $E^{\pm}=\int |{\bf z}^\pm|^2 \, d^3 x=\int e^{\pm}({\bf k}) \,d^3 k$. When the energies are supplied to the system at large scales, they get redistributed over scales by nonlinear interactions, and removed from the system at small dissipation scales. One can argue  that the energy gets redistributed predominantly over the modes whose wevevectors are approximately normal to the strong guide magnetic field. We will concentrate on the so-called balanced case, when $e^+\sim e^-$, and we can therefore represent the Fourier  energy spectra in the form 
\begin{eqnarray}
e^\pm(k_\|, k_\perp)=f^\pm(k_\|/k_\perp)k_\perp^{-\alpha},
\label{e_elsasser}
\end{eqnarray}
where $f^\pm(x)$ do not vary significantly for $x<C$ and decline fast for $x>C$, with some constant $C$. Here $k_\|$ is the wavevector in the direction of the uniform field ${\bf B}_0$, and ${\bf k}_\perp$ is the wavevector in the field-perpendicular direction. This form of the spectral functions is motivated by the fact that the energy redistribution occurring due to small-scale fluctuations (large $k$) is predominantly normal to the direction of the {\em local} guide field, which is the field produced by large-scale fluctuations. Therefore, compared to the direction of the {\em global} uniform field, the energy spectrum is smeared inside the small angle $\theta_0\sim b_{rms}/B_0$, which implies a wedge-shaped energy-containing domain $k_\|< \theta_0 k_\perp$, or, the spectral function (\ref{e_elsasser}) with $C\sim\theta_0$, \citep[e.g.,][]{cho_v00,maron_g01,chen10}. One can then write down a model equation for the spectral function (\ref{e_elsasser}), using certain closure assumptions. Differentiating $e^{\pm}({\bf k})$ with respect to time, iterating equation (\ref{mhd-elsasser}) once, and splitting the forth-order correlation functions of ${\bf z}$'s into pair-wise correlations using Gaussian rule, we get\footnote{Some extra assumptions are made in obtaining this equation, for instance it is assumed that the cross-correlation $\langle {\bf z}^+\cdot {\bf z}^- \rangle$ is absent, see, e.g., \cite{goldreich_s95}. It should however be borne in mind that this equation is not rigorously derived from~(\ref{mhd-elsasser}); it should be considered only as a model equation or as a plausible two-point closure.}   
\begin{eqnarray}
{\partial_t}e^{\pm}(k_\|, k_\perp)=\int M({\bf k},{\bf k}_1,{\bf k}_2)\Theta^{\pm}(k_{2\|},k_{2\perp})e^{\mp}(k_{2\|},k_{2\perp})\left[ e^{\pm}(k_{1\|},k_{1\perp})-e^{\pm}(k_{\|},k_\perp) \right] \times \nonumber \\
\times \delta (k_\|-k_{1\|}-k_{2\|})\delta({\bf k}_\perp-{\bf k}_{1\perp}-{\bf k}_{2\perp})\, d^3\, k_1\, d^3\, k_2.
\label{closure}
\end{eqnarray}
In this equation, the kernel has the form  
%\begin{eqnarray}
$M_{{\bf k},{\bf k}_1,{\bf k}_2} \propto ({\bf k}_{\perp}\times {\bf k}_{2 \perp})^2({\bf k}_{\perp}\cdot {\bf k}_{1\perp})^2/(k_{\perp}^2k_{1\perp}^2k_{2\perp}^2)$, 
%\label{kernel}
%\end{eqnarray}
and the $\Theta^{\pm}$ functions depend on the assumptions about the nonlinear interaction made in the model. In general we argue that these functions should concentrate in the region where the nonlinear interaction is essential, and inside this region they should scale as the inverse time of nonlinear interaction, that is, $\Theta^\pm(k_\|,k_\perp)\propto 1/\tau(k_\perp)$. This can be summarized as follows,
$\Theta^{\pm}(k_\|, k_\perp)=g^{\pm}(k_\|, k_\perp)k_\perp^{-\delta}$, 
where $g^\pm(k_\|,k_\perp)\approx {\rm const}$ in the region of nonlinear interaction, and the nonlinear interaction time scales as $\tau(k_\perp)\propto k_\perp^{\delta}$. 

To understand better our model (\ref{closure}), consider particular examples. In the case of weak turbulence, $g(k_\|,k_\perp)\approx {\rm const}$ in a quite narrow region  $k_\|V_A\leq 1/\tau(k_\perp)$ compared with the $k_\|$-widths of the functions $e^\pm$, and it declines fast for $k_\|V_A \geq  1/\tau(k_\perp)$. Therefore, $\Theta(k_\|,k_\perp)$ is a broadened delta-function of $k_\|$, obeying $\int \Theta(k_\|,k_\perp)dk_\|={\rm const}$. One then recovers the theory by \cite{galtier_nnp00}. In the case of strong turbulence, one expects the nonlinear interaction to be important in the same region where the energies $e^{\pm}$ are concentrated, that is, $g^{\pm}\approx {\rm const}$ in the region $k_\| \leq \theta_0k_\perp$, it declines outside of this region, and $\tau \sim \lambda/z(\lambda) $. This way we recover the \cite{goldreich_s95} theory. If in addition to the assumptions of the GS theory, one assumes that there is persistent dynamic angular alignment between magnetic and velocity fluctuations, which reduces the nonlinear interaction by $\theta_\lambda\sim \lambda^{1/4}$, one needs to multiply the kernel in (\ref{closure}) by $\theta_\lambda^2\sim k_2^{-1/2}$, and assume that $\tau \sim \lambda /(v_\lambda \theta_\lambda )$. One then recovers the theory by \cite{boldyrev06}. In view of this, we stress that model (\ref{closure}) provides a useful description of the spectral energies in MHD turbulence, however, 
%it does not allow one to derive the spectrum in any rigorous fashion, since 
it crucially depends on the  scaling assumptions about the interaction time, incorporated in the model.\footnote{The same statement is true for the so-called EDQNM closures often used to derive equations of type (\ref{closure}) for the spectra of strong turbulence. While providing physically reasonable models of turbulence, such equations are not derived from first principles and they crucially depend of model assumptions.} 

The steady spectrum of turbulence can then be found by requiring that the collision integral in the rhs of (\ref{closure}) is zero. This leads to the spectrum of weak turbulence \citep{ng_b96,galtier_nnp00}:  
%\begin{eqnarray}
$E^{\pm}(k_\|,k_\perp)=e^{\pm}(k_\|,k_\perp)2\pi k_\perp\propto f^\pm_w(k_\|)k_\perp^{-2}$,
%\label{galtier}
%\end{eqnarray}
where $f^\pm_w(k_\|)$  depend on the details of large-scale driving. The field-perpendicular spectrum of strong turbulence in \cite{goldreich_s95} theory is then found as 
%\begin{eqnarray}
$E^{\pm}(k_\perp)=\int e^{\pm}(k_\|,k_\perp)2\pi k_\perp \, dk_\| \propto k_\perp^{-5/3}$,
%\label{gs}
%\end{eqnarray}
while the spectrum in \cite{boldyrev06} theory is 
%\begin{eqnarray}
$E^{\pm}(k_\perp)=\int e^{\pm}(k_\|,k_\perp)2\pi k_\perp \, dk_\| \propto k_\perp^{-3/2}$. 
%\label{b}
%\end{eqnarray}
Numerical simulations do produce the spectrum $k_\perp^{-2}$ for weak turbulence \citep{perez_b08}, and the spectrum $k_\perp^{-3/2}$ for strong turbulence with a strong guide field ${\bf B}_0$, e.g., \citep{muller_g05,mason_cb08}.

\section{The spectrum of the residual energy}
Recently, it has been realized that significant role in turbulence dynamics is played by the so-called residual energy, that is, the energy difference between magnetic and kinetic fluctuations, $E_r=E_v-E_b$, see \citep{boldyrev_p09,wang11,boldyrev_pbp11}. Indeed, the complete  description of the second-order statistics of two fluctuating  fields, ${\bf v}$ and ${\bf b}$  requires {\em three} independent correlation functions. Two of them are provided by the autocorrelation functions of the Elsasser variables, that is, the energy spectra (\ref{e_elsasser}). The third one, the cross-correlation function,  is the residual energy 
%\begin{eqnarray} 
$E_r=\int ({\bf z}^+\cdot {\bf z}^-)\, d^3 x=\int Re[ e^r({\bf k})]\, d^3 k$, 
%\label{residual}
%\end{eqnarray}
where, by definition, $e^r({\bf k})={\bf z}^+({\bf k})\cdot {\bf z}^{-*}({\bf k})$. It is easy to see that $e^r({\bf k})$ is a complex function, while the residual energy spectrum is its real part. 

The residual energy has been previously addressed in the literature \citep[e.g.,][]{pouquet_fl76,grappin_pl83,zank_ms96,muller_g05,ng_b07,chen11a}, but it has been studied to a much lesser extent compared to the Elsasser energies, possibly because it is not a conserved quantity, it is not sign-definite, and and it does not exhibit a cascade in a turbulent regime. As a result, it cannot be expressed through the conserved Elsasser energies (or, equivalently, through the total energy or cross-helicity), which are commonly used in theoretical models and measured in observations. In fact, in many studies of MHD turbulence the residual energy is explicitly or implicitly assumed to be zero, see e.g., \citep{galtier_nnp00}. In this section we propose a model for the residual energy, analogous to Eq.~(\ref{closure}). We demonstrate that in contrast with the scaling of the Elsasser energies, the scaling of the residual energy is quite robust, that is, it depends to a mush lesser extent on the arbitrary scaling assumptions incorporated in the model. 

To obtain the equation for the residual energy, we first note that in the absence of the nonlinear interaction, the spectral residual energy $e^r({\bf k})={\bf z}^+({\bf k})\cdot {\bf z}^{-*}({\bf k})$ oscillates in time, since ${\bf z}^+({\bf k})\propto \exp(ik_\|v_A t)$ and ${\bf z}^-({\bf k})\propto \exp(-ik_\|v_At)$. When the nonlinear interaction is present, the residual-energy evolution equation should contain the terms describing  interaction of the residual energy with the Elsasser fields, $\sim e^r e^{\pm}$, and  generation of the residual energy by the Elsasser fields, $\sim e^{+}e^{-}$. It has been recently realized that the terms describing the generation of the residual energy by the Elsasser fields are essentially nonzero \citep{wang11}, meaning that residual energy is spontaneously generated by turbulent dynamics even if it is zero initially. We start our discussion of the residual energy with more detailed consideration of these terms.  

It is crucial to note that the terms describing generation of the residual energy by the Elsasser fields should have the same {\em dimension} as the rhs of Eq.~(\ref{closure}). Indeed, the residual energy has the same dimension as the Elsasser energies, and it is generated due to same nonlinear interactions.  We however do not need the exact structure of this term, rather, we need to know its {\em scaling} with respect to the wavenumber. It turns out that this scaling is rather universal. Indeed, the term in the rhs of Eq.~(\ref{closure}) describes constant flux $J$ of the Elsasser energies $e^{\pm}({\bf k}_\perp)$ in the field-perpendicular direction, that is,  it should scale as $\frac{1}{k_\perp}\frac{\partial}{\partial k_\perp} J\propto k_\perp^{-2}$, no matter what particular model of turbulence is assumed. The term describing generation of the residual energy should then have the same scaling, although its structure is different. We therefore model the residual-energy generating term as $\alpha(k_\|, k_\perp)$ with the only requirement that it is concentrated in the region where the Elsasser energies are concentrated, and it obeys $\int \alpha(k_\|, k_\perp)\, dk_\| \propto k_\perp^{-2}$. 

The term describing relaxation of the residual energy due to its interaction with the Elsasser energies can be generally modeled as $-\gamma(k_\|, k_\perp)e^r(k_\|, k_\perp)$, where the relaxation rate $\gamma$ depends on the spectrum of the Elsasser fields, and it is concentrated in the region where the nonlinear interaction is present. We now collect all the three terms to formulate our model equation for the residual energy:
\begin{eqnarray}
\partial_t e^r(k_\|, k_\perp)=2ik_\|v_A e^r - \gamma(k_\|, k_\perp)e^r+ \alpha(k_\|, k_\perp). 
\label{model}
\end{eqnarray}

\section{Discussion}
We now apply our formalism to the cases of weak and strong turbulence. For weak turbulence, the energy cascades predominantly in the field-perpendicular direction for each $k_\|$, so that the field-parallel structure of the spectrum does not change with $k_\perp$. Moreover, in this case the residual-energy generating term can be shown to be negative \citep{wang11}. We can therefore write (restoring the dimensional coefficients) that $\alpha(k_\|, k_\perp)=-\alpha_w (v^4_{rms}/v_A) k_\perp^{-2}$ for $k_\|<k_{0\|}$, where $\alpha_w$ is a dimensionless constant and $k_{0\|}\sim 1/L_\|$ is the field-parallel spectral width of the Elsasser fields. Weak turbulence theory also allows one to estimate the time of nonlinear interaction of the fields, which gives
$\gamma=\beta (v^2_{rms}/v_A) k_\perp$ for $k_\|\approx 0$ (it will be clear momentarily why only the region $k_\|\approx 0$ is relevant here), and $\beta$ is a dimensionless constant. Solving  Eq.~(\ref{model}) for long times, we get:
\begin{eqnarray}
e_v-e_b=Re [ e^r(k_\|,k_\perp)]= -\frac{\alpha_w \beta v^2_{rms}\epsilon^4 k^{-1}_\perp}{\beta^2 k^2_\perp \epsilon^4 +4k^2_\|},
\label{er}
\end{eqnarray}
where $\epsilon=v_{rms}/v_A\sim b_{rms}/B_0\ll 1$. The residual energy is concentrated in a narrow region around $k_\| \leq \beta \epsilon^2 k_\perp /2$, in agreement with previous findings \citep{boldyrev_p09,wang11}. The phase-volume compensated field-perpendicular spectrum of the residual energy then has the structure
\begin{eqnarray}
E_r(k_\|,k_\perp)=Re [e^r(k_\|,k_\perp)]2\pi k_\perp =-f_w(k_\|/k_\perp)k_{\perp}^{-2},
\label{er_weak1}
\end{eqnarray}
where $f_w(x)=v_{rms}^2 \alpha_w\beta  \epsilon^4/(\beta^2 \epsilon^4+4x^2)$, as follows from~(\ref{er}). We can also define the field-perpendicular spectrum
\begin{eqnarray}
E_r(k_\perp)=\int E_r(k_\|,k_\perp)\,dk_\|=\alpha_w \pi v^2_{rms}\epsilon^2 k_\perp^{-1}\sim - v_{rms}^2 \epsilon^2 k_\perp^{-1}.
\label{er_weak2}
\end{eqnarray}
For the case of strong turbulence, the spectra of $e^\pm(k_\|, k_\perp)$ are concentrated in the region $k_\|\leq \theta_0k_\perp$, and it is reasonable to assume that the function  $\alpha(k_\|, k_\perp)$ is concentrated in the same region.  Restoring the dimensional parameters, one can therefore write $\alpha(k_\|,k_\perp)= -\alpha_s (v^3_{rms}/L_\perp)(\theta_0k_\perp)^{-1}k_\perp^{-2}$ for a given $k_\|$ inside the region $k_\|\leq \theta_0k_\perp$, where $\alpha_s$ is a dimensionless parameter, and $L_\perp$ is the integral field-perpendicular scale of the fluctuations. Note that the power of $k_\perp$ is fixed by the requirement $\int \alpha(k_\|, k_\perp)\, dk_\|\propto k_\perp^{-2}$. One can also assume the power-law behavior for the relaxation rate, $\gamma(k_\|, k_\perp) = \gamma k_\perp^{\mu}$ within the same region, where $\gamma$ is a dimensional parameter. It should be noted, however, that in contrast with the function $\alpha(k_\|, k_\perp)$, whose scaling could be established on dimensional grounds, the scaling of the function $\gamma(k_\|,k_\perp)$ cannot be easily derived. One can only argue that this relaxation rate should compete with the linear frequency only in the region where the Elsasser energies themselves are concentrated, that is, $\mu\leq 1$.  It is interesting, however, that this bound is enough to establish the field-perpendicular spectrum of the residual energy.  The solution of (\ref{model}) takes the form
\begin{eqnarray} 
e_v-e_b=Re [e^r(k)]=\frac{\gamma k_\perp^{\mu} }{\gamma^2 k_\perp^{2\mu} +4k^2_\|v_A^2}\,\alpha(k_\|, k_\perp).
\end{eqnarray}
To find the field-perpendicular energy spectrum, one integrates this result over $k_\|$. One does not need however to integrate the function $\alpha(k_\|, k_\perp)$, since the prefactor is a narrower function (a broadened $\delta$-function, in fact). The integral over $k_\|$ is then independent of $\gamma k^\mu$, and the result is:
\begin{eqnarray}
E_r(k_\perp)=\int Re [e_r(k_\|,k_\perp)] 2\pi k_\perp \,dk_\|\sim -v^2_{rms}L_\perp^{-1}k_\perp^{-2},
\label{er_strong1}
\end{eqnarray}
where we used $\theta_0\sim v_{rms}/v_A$. This result is in agreement with numerical studies \citep[e.g.,][]{muller_g05}. It is also of interest to establish the value of $\mu$. This can be inferred from numerical simulations if one evaluates $E_r(k_\|=0, k_\perp)\propto k_\perp^{-2-\mu}$. Our simulations (that will be reported elsewhere) indicate that, quite interestingly,  $\mu\approx 1$, which allows us to write the general expression for the residual energy in the form:
\begin{eqnarray}
E_r(k_\|,k_\perp)=Re [e^r(k_\|,k_\perp)]2\pi k_\perp =-f_s(k_\|/k_\perp)k_{\perp}^{-3},
\label{er_strong2}
\end{eqnarray} 
where $f_s(x)\approx 1/\theta_0$ for $x<\theta_0$, and it declines for $x>\theta_0$. Relations (\ref{model}), (\ref{er_weak1}-\ref{er_weak2}), and (\ref{er_strong1}-\ref{er_strong2}) are the main results of our work; they provide a model for residual energy observed in the solar wind and in numerical simulations. 

\acknowledgments 
This work was supported   
by the US DOE Awards DE-FG02-07ER54932, DE-SC0003888, DE-SC0001794, the NSF Grant PHY-0903872, the NSF/DOE Grant AGS-1003451, and by the NSF Center for Magnetic Self-organization in Laboratory and Astrophysical Plasmas at the University of Wisconsin-Madison.

%\bibliographystyle{asp2010}
%\bibliography{References_stas}

\end{document}